\documentstyle[aps,12pt]{revtex}

\begin{document}
\draft
\author{O.B.Zaslavskii}
\address{Department of Physics, Kharkov State University, Svobody Sq.4, Kharkov,\\
310077, UKRAINE\\
e-mail: oleg.b.zaslavskii@univer.kharkov.ua}
\title{Geometry of nonextreme black holes near the extreme state}
\maketitle

\begin{abstract}
Nonextreme black hole in a cavity can achieve the extreme state with a zero
surface gravity at a finite temperature on a boundary, the proper distance
between the boundary and the horizon being finite. The classical geometry in
this state is found explicitly for four-dimensional spherically-symmetrical
and $2+1$ rotating holes. In the first case the limiting geometry depends
only on one scale factor and the {\it whole }Euclidean manifold is described
by the Bertotti-Robinson spacetime. The general structure of a metric in the
limit under consideration is also found with quantum corrections taken into
account. Its angular part represents a two-sphere of a constant radius. In
all cases the Lorentzian counterparts of the metrics are free from
singularities.
\end{abstract}




One of the most intriguing issues in black hole physics is connected with
the nature of the extreme state. In particular, it concerns the possibility
of a thermodynamic description of such objects which is highly nontrivial
because the Hawking temperature $T_{H}$ in the state under discussion. In
Ref. \cite{[1]} it was suggested to consider such objects as having an
arbitrary temperature $T_{0}$ measured at infinity which determines the
period in Euclidean time. This proposal is motivated by the qualitative
difference between the topology of extreme and non-extreme black holes: the
Euclidean manifold of an extreme one is regular irrespective of the value $%
T_{0}$ in contrast with non-extreme holes for which the absence of conical
singularities demands $T_{0}=T_{H}$ Reasonings connected with topology \cite
{[1]}, \cite{[2]} show that one should ascribe the entropy $S=0$ instead of
the Bekenstein-Hawking value $S=A/4$ ($A$ is the surface area of a horizon).

However, further investigations showed that the deviation of $T_{0}$ from
its Hawking value $T_{H}=0$ entails unavoidable divergences in the
stress-energy tensor of quantum fields either for two-dimensional \cite{[3]}
or four-dimensional \cite{[4]} extreme black holes and for this reason the
possibility of $T_{H}\neq T_{H}$ is unacceptable physically. One can say
that the principle of thermal equilibrium is more fundamental that the
requirement of the regularity for the Euclidean manifold. The latter turns
out to be simply irrelevant for extreme fields in a corresponding
background. In the case $T_{0}=T_{H}$ the difficulties indicated above are
removed but the possibility to build finite temperature description of
extreme holes is lost.

Meanwhile, recently it was traced \cite{[5]} how a black hole can approach
the extreme state in the topological sector of nonextreme configurations
with $S=A/4$ when for the Reissner-Nordstr\"{o}m (RN) metric $m\rightarrow e$
($m$ is a mass, $e$ is a charge). In so doing, the thermodynamic equilibrium
is fulfilled at every stage of the limiting transition, so difficulties due
to $T_{0}\neq T_{H\,}$ cannot arise at all. In the grand canonical ensemble
approach when a system is characterized by the local temperature $T=\beta
^{-1}$ and potential $\Phi \,$at the boundary $r_{B}$ it turns put that
among all boundary data $(\beta ,r_{B},\Phi )$ there exists such a subset $%
(\beta ,r_{B},\Phi (\beta ,r_{B}))$ for which (i) the extreme state is
realized at finite temperature, (ii) the horizon has the same area as the
boundary surface but is situated at proper distance $L$ from it (unlike the
extreme topological sector where $L=\infty ).$ In some sense, a nonextreme
hole imitates the extreme one, so $T_{0}\rightarrow 0$ (but $T$ is finite).

The geometrical properties indicated above are rather unexpected and needed
to be clarified. In this paper I show that for a wide class of black holes
there exists the extreme state of non-extreme black holes which has the
universal form of the limiting metric and find it explicitly. More exactly,
I consider (i) spherically-symmetrical configurations near the extreme
state; (ii) ''the ultra-extreme case'' and (iii) 2+1-dimensional rotating
holes. It is remarkable that in spite of the variety of types of black hole
solutions and their parameters shortening of description occurs (as we will
see) near the state in question.

Consider the Euclidean black hole metric 
\begin{equation}
ds^{2}=f(r)d\tau ^{2}+f^{-1}dr^{2}+r^{2}d\omega ^{2},\text{ }d\omega
^{2}=d\theta ^{2}+d\phi ^{2}\sin ^{2}\theta  \label{(1)}
\end{equation}
The Euclidean time takes its values in the range $0\leq \tau \leq
T_{0}^{-1}=T_{H}^{-1}.$ Let us introduce the new variable $\varphi =2\pi
T_{0}\tau $ where now $0\leq \varphi \leq 2\pi .$ Then 
\begin{equation}
ds^{2}=\left( \frac{\beta }{2\pi }\right) ^{2}d\varphi
^{2}+dl^{2}+r^{2}d\omega ^{2},  \label{(2)}
\end{equation}
$\beta (r(l))=\beta _{0}[f(r)]^{1/2}$ is the inverse local temperature at an
arbitrary point $r_{+}\leq r\leq r_{B},$ $r_{+}$ is the radius of the event
horizon, $l$ is the proper distance between $r_{+}$ and $r.$

For the spacetime \ref{(1)}, \ref{(2)} the equilibrium condition reads 
\begin{equation}
\beta =\beta _{0}[f(r_{B})]^{1/2},\text{ }T_{0}=T_{H}=\frac{f^{\prime
}(r_{+})}{4\pi },  \label{(3)}
\end{equation}
where a prime denotes the derivative with respect to a corresponding
argument, $\beta $ is taken at the boundary $r_{B}.$

For the nearly extreme state $\beta _{0}\rightarrow \infty $ but there
exists the subset of boundary data for which $r_{+}\rightarrow r_{B}$ in
such a way that the limit can be reached at finite $\beta $ with the finite
distance $l=l_{B}$ between $r_{+}$ and $r_{B}$ \cite{[5]}. In so doing, the
coordinate $r$ becomes ill-defined and it is more convenient to use $l$ or
its dimensionless analog as a new coordinate. Let us choose this coordinate
according to 
\begin{equation}
r-r_{+}=4\pi T_{0}b^{-1}(\frac{\sinh x}{2})^{2},\text{ }b=\frac{f^{\prime
\prime }(r_{+})}{2}  \label{(4)}
\end{equation}
As in the limit in question the region $r_{+}\leq r\leq r_{B}$ shrinks we
can expand $f(r)$ in a power series $f(r)=4\pi
T_{0}(r-r_{+})+b(r-r_{+})^{2}+...$ near $r=r_{+}.$ Then after substitution
into \ref{(2)}, \ref{(3)} we obtain 
\begin{equation}
ds^{2}=b^{-1}(d\varphi ^{2}\sinh ^{2}x+dx^{2})+r_{B}^{2}d\omega ^{2},\text{ }%
x=lb^{1/2}  \label{(5)}
\end{equation}
For the RN hole $f=(1-r_{+}/r)(1-r_{-}/r)$ and in the extreme limit $%
r_{+}=r_{-}=r_{B},$ $b=r_{B}^{-2}.$ Then 
\begin{equation}
ds^{2}=r_{B}^{2}(d\varphi ^{2}\sinh ^{2}x+dx^{2}+d\omega ^{2})  \label{(6)}
\end{equation}
(Euclidean version) or 
\begin{equation}
ds^{2}=r_{B}^{2}(-dt^{2}\sinh ^{2}x+dx^{2}+d\omega ^{2})  \label{(7)}
\end{equation}
(Lorentzian version).

This is nothing more than the Bertotti-Robinson (BR) spacetime \cite{[6]}, 
\cite{[7]}.

That the BR metric is relevant for the description of the extreme RN hole
was already pointed out in the literature \cite{[4]}, \cite{[8]}, \cite{[9]}
but in quite different context when the extreme RN hole was considered in
the topological sector corresponding to extreme black holes, $%
f=(1-r_{+}/r)^{2}$ from the very beginning. Then, expanding the metric
coefficients near $r=r_{+}$ and introducing the new radial variable
according to $r-r_{+}=r_{B}\rho ^{-1}$one obtains 
\begin{equation}
ds^{2}=r_{B}^{2}\rho ^{-2}(-dt_{1}^{2}+d\rho ^{2}+\rho ^{2}d\omega ^{2},
\label{(8)}
\end{equation}

Two metrics \ref{(7)} and \ref{(8)} look differently but in fact are
equivalent locally. This can be seen from the representation of the BR
spacetime as the metric on the hyperboloid \cite{[7]}. Putting $%
w^{2}+v^{2}-u^{2}=1,$ $ds^{2}=du^{2}-dv^{2}-dw^{2}$ and making substitution
for the part of the hyperboloid $u\geq 0$%
\begin{equation}
u=\cosh t\sinh x,\text{ }v=\sinh t\sinh w,\text{ }w=\cosh x  \label{(9)}
\end{equation}
we obtain eq.\ref{(7)}. Another substitution 
\begin{equation}
u=(2\rho )^{-1}(t_{1}^{2}-\rho ^{2}+1),\text{ }v=(2\rho
)^{-1}(t_{1}^{2}-\rho ^{2}-1),\text{ }w=\rho ^{-1}t_{1}  \label{(10)}
\end{equation}
(for simplicity we put $r_{B}=1$ for a moment) gives us eq. \ref{(8)}. Here
two coordinate systems are connected by formulas 
\begin{equation}
t_{1}/r_{B}=e^{t}\coth x,\text{ }\rho /r_{B}=e^{t}(\sinh x)^{-1},
\label{(11)}
\end{equation}
\begin{equation}
\cosh x=t_{1}/\rho ,\text{ }e^{2t}=(t_{1}^{2}-\rho ^{2})/r_{B}^{2}
\label{(12)}
\end{equation}

I stress that whereas the BR metric \ref{(8)} gives only the approximate
representation of the RN one of an extreme hole in a small vicinity of $%
r=r_{+}$ solution \ref{(6)} refers to the {\it whole }Euclidean manifold
whose four-volume in our case is finite. It is clear from Eqs.\ref{(6)}, \ref
{(7)} that the proper distance between a horizon $x=0$ and any other point $%
x>0$ is finite whereas the distance between $\rho =\infty $ and any other $%
\rho <\infty $ is infinite. It is this property which gives rise to
qualitatively distinction between topologies of extreme and nonextreme holes
and their entropies \cite{[1]}, \cite{[2]}.

Thus, starting from the original RN metric we arrive in the limit $%
m\rightarrow e$ at two different versions of the BR spacetime corresponding
to different Killing vectors $\partial /\partial t$ and $\partial /\partial
t_{1}$ depending on a topological sector \cite{[10]}. That it is eq. \ref
{(7)} (but not eq. \ref{(8)}) which is relevant for our problem is not
incidental. The Hawking effect is intimately connected with the existence of
a horizon which makes the outer region geodesically incomplete. Either Eq. 
\ref{(7)} or Eq.\ref{(8)} possesses this property but \ref{(7)} is ''more
incomplete'' in the sense that only the part $t_{1}>\rho $ of \ref{(8)} is
mapped onto \ref{(7)} as follows from \ref{(11)}, \ref{(12)}. As a result, $%
T_{H}=(2\pi )^{-1}$ for \ref{(7)} in the accordance with the finite boundary
temperature in our problem whereas $T_{B}=0$ for \ref{(8)}: roughly
speaking, the more information lost, the more there is a temperature.

The BR metric is obtained above from the RN one and for this reason
represents only classical geometry. Quantum effects will certainly change
its form. (In particular, Riemann curvature $R\neq 0$ in general while $R=0$
for the BR spacetime). It is remarkable, however, that {\it all} effects of
back reaction for the model \ref{(1)} are encoded in the coefficients $b$
only, so the metric is described by \ref{(5)} instead of \ref{(6)} where $b$
takes its RN value $r_{B}^{-2}.$ It is worth also paying attention that the
explicit form of boundary data for which the state in question is achieved ($%
\beta =4\pi r_{B}\Phi (1-\Phi ^{2})^{-1}$ for the classical RN hole \ref{(5)}%
) was not used in derivation at all as well as the form of field equations.
Such universality is explained by that we consider the state for which {\it %
by definition} $T_{0}\rightarrow 0.$ Quantum effects can change the
connection between $T_{H}$ and black hole parameters as well as relationship
between boundary data (say, between, $\beta $ and $\Phi )$ for a ''dressed''
hole as compared with a bare one. Nonetheless, if a theory admits the
existence of an extreme state in the above sense the limiting form of the
metric with finite boundary conditions is described by Eq.\ref{(5)}.

Bearing in mind the possible role of quantum effects, of interest is a
metric more general that \ref{(1)}: 
\begin{equation}
ds^{2}=U(r)d\tau ^{2}+V^{-1}(r)dr^{2}+r^{2}d\omega ^{2}  \label{(13)}
\end{equation}

Then, repeating all manipulations step by step we arrive at 
\begin{equation}
ds^{2}=b^{-1}(ad\varphi ^{2}+dx^{2})+r_{B}^{2}d\omega ^{2}  \label{(14)}
\end{equation}
Here $a=4\sinh ^{2}x/2(1+\cosh ^{2}x/2),$ $b=V^{\prime \prime }(r_{+})/2,$ $%
c=V^{\prime }(r_{+})U^{\prime \prime }(r_{+})/V^{\prime \prime
}(r_{+})U^{\prime }(r_{+}).$

In general, if $c\neq 1$ the Riemann curvature is no longer constant but
depends on a point due to effects of back reaction.

The developed approach is also applicable to two-dimensional metrics which
can be obtained by discarding the term $d\omega ^{2}.$ It is known that in
the topological sector of extreme holes there arise weak divergences of the
stress-energy tensor of quantum fields on the event horizon discussed
recently in the context of two-dimensional dilaton gravity \cite{[11]}.
These divergences are due to the property $T_{H}=0$ \cite{[4]} and are
absent in our case where $T_{H}\neq 0$ for the limiting form of the metric 
\ref{(6)} and \cite{(14)}.

Let us discuss briefly the case of so-called ultraextreme black holes for
which, in the extreme state, $b=0$ by definition \cite{[12]}. One can check
that finite boundary conditions are again possible due to $r_{+}\rightarrow
r_{B}$ in spite of $T_{H}\rightarrow 0.$ In this case the substitution $%
r-r_{+}=\pi T_{H}l^{2}$ shows that terms of the third order and higher in
the expansion of $f(r)$ near $r=r_{+}$ are negligible and 
\begin{equation}
ds^{2}=l^{2}d\varphi ^{2}+dl^{2}+r_{B}^{2}d\omega ^{2}.  \label{(15)}
\end{equation}

In other words, we obtain the direct product of the two-dimensional Rindler
space and a two-sphere of a constant radius.

I stress that this result is by no means the trivial consequence of the
known fact that the metric of a generic black hole can be represented by the
Rindler one near the horizon. Such representation is approximate, valid only
in a small region near the event horizon, applicable to a nonextreme black
hole and has nothing to do with the thermodynamic approach we deal with.
Meanwhile, our black hole is ultraextreme and the four-dimensional volume of
an Euclidean manifold is finite. The metric \ref{(15)} appeared as a result
of the limiting transition from the nonextreme state when $f^{\prime }(r_{+})
$ is small but nonzero while $f^{\prime \prime }(r_{+})=0$ by definition;
had we started from the metric with $f=const(r-r_{+})^{3}$ at once we would
have obtained the metric which has nothing to do with Eq.\ref{(15)}.

Now we will analyze the case of $2+1$ black holes \cite{[13]} which is
achieved due to rotation. The metric reads \cite{[13]} 
\begin{eqnarray}
ds^{2} &=&-N^{2}dt^{2}+N^{-2}dr^{2}+r^{2}(N^{\phi }dt+d\phi ^{2})^{2},\text{ 
}  \label{(16)} \\
N^{2} &=&-M+(r/l)^{2}+J^{2}/4r^{2},\text{ }N^{\phi }=-J/2r^{2}  \nonumber
\end{eqnarray}

As we will be interested in the metric near the horizon $r_{+}$ it is
convenient to redefine the angular variable according to $\phi \rightarrow
\phi -N^{\phi }(r_{+})t.$ Then 
\begin{equation}
ds^{2}=-N^{2}dt^{2}+N^{-2}dr^{2}+r^{2}(N^{\phi }dt+d\phi )^{2},\text{ }%
N^{\phi }=J(\frac{1}{2r_{+}^{2}}-\frac{1}{2r^{2}})  \label{(17)}
\end{equation}

Following the general approach for finite-size thermodynamics of $2+1$ holes
(see \cite{[14]}, \cite{[15]} and references therein) let us consider the
grand canonical ensemble for which the set of boundary conditions includes
in the case under discussion $r_{B},$ $\beta $ and the angular velocity $%
\Omega $ of the heat bath with respect to zero angular momentum observers: 
\begin{equation}
\Omega =JN^{-1}(1/2r_{+}^{2}-1/2r^{2}),  \label{(18)}
\end{equation}

\begin{equation}
\beta =\beta _{0}N,\text{ }\beta _{0}=2\pi r_{+}(M^{2}-J^{2}/l^{2})^{-1/2}.
\label{(19)}
\end{equation}

It is convenient to introduce dimensionless quantities $y=r_{+}/r_{B},$ $%
q=\Omega r,$ $\sigma =\beta /2\pi l,$ $z=J/Ml.$ Then after some algebraic
manipulations one obtains from \ref{(18)}, \ref{(19)}: 
\begin{eqnarray}
y &=&\{(1-q^{2})[1+\sigma ^{2}(1-q^{2})]\}^{-1/2},\text{ }  \label{20} \\
z &=&2\sigma q[q^{2}+\sigma ^{2}(1-q^{4})]^{-1}\{(1-q^{2})[1+\sigma
^{2}(1-q^{2})]\}^{1/2}  \nonumber
\end{eqnarray}

We are interested in the possibility of finding the finite-temperature
solutions of these equations for the extreme state ($\sigma <\infty ,$ $%
z=1). $ It is seen from \ref{20} directly that such a solution does exist if 
$y=1$ ($r_{+}=r_{B}$ as well as in the $3+1$ case) and boundary data are
restricted by the condition 
\begin{equation}
\sigma =q/(1-q^{2})  \label{(21)}
\end{equation}

The thermodynamic description needs the transition from the Lorentzian
picture to the Euclidean one that for rotating holes gives rise to the
complexification of a metric \cite{[16]}. However, to avoid subtleties which
are irrelevant for the issue under consideration, I will list a metric at
once in the Lorentzian form. After expanding $N^{2}$ and $N^{\phi }$ in a
power series near $r=r_{+}$ and using the substitution \ref{(4)} one can
obtain 
\begin{equation}
ds^{2}=r_{B}^{2}[-dt_{1}^{2}\sinh ^{2}x+dx^{2}+(d\phi +\Omega
_{1}dt_{1})^{2}],\text{ }\Omega _{1}=l(\sinh ^{2}x/2)/4r_{B}  \label{(22)}
\end{equation}
where time is normalized according to $t_{1}=2\pi T_{0}t.$ Now only two
dimensionless parameters are independent among all boundary data - say $%
l/r_{B}$ and $q$ whereas $\sigma $ is determined by Eq.\ref{(21)}.

Thus, we obtained the generalization of the BR metric to the case of
rotation as the limiting form of $2+1$ holes in the state under discussion.

It is worth noting that if $\sigma \rightarrow 0$ ($T\rightarrow \infty $)
we obtain the extreme state not simply at finite temperature but even in the
high-temperature limit. In so doing, $q\rightarrow 0$ according to \ref{(21)}%
, so the extreme hole $(J=Ml)$ is slowly rotating.

To summarize, it turned out that in the topological sector of non-extreme
black holes a metric takes the universal form near the extreme state. Its
classical geometry is described by a single parameter $r_{B}$ which enters
the metric simply as a scale factor in the static case and by two parameters
for rotating $2+1$ holes. The essential feature of the obtained solutions is
the absence of singularities of spacetime in contrast to the truly
(topologically) extreme black holes for which singularities exist behind a
horizon. Although in the thermodynamic approach only the region $r>r_{+}$ is
relevant it is possible that the above results shed light on the issue of
singularities. Indeed, due to property chosen boundary data all the range of
a radial coordinate shrinks to the point $r=r_{+}$ in such a way that
Lorentzian versions of metrics are free from singularities. In this sense it
is thermodynamics which is responsible for removing singularities in the
Lorentzian versions of metrics. Perhaps, nature forbids the existence of
truly extreme black holes as thermodynamic objects (see the beginning of the
paper) but gives us instead the hint to another possible approach to the
problems of either the extreme state or singularities in the framework of
gravitational thermodynamics.

Of interest is to generalize the obtained results to non-spherical and
rotating four-dimensional black holes.


%
%

%
%

\end{document}